\documentclass[usenatbib,usegraphicx]{mn2e}

\usepackage{amsmath}
\usepackage{psfig}
\usepackage{color}

\def\apj{ApJ}
\def\apjl{ApJL}
\def\apjs{ApJS}
\def\aap{A\&A}
\def\mnras{MNRAS}
\def\aj{AJ}
\def\nat{Nature}
\def\araa{ARA\&A}
\def\ssr{Space Sci. Rev.}
\def\pasj{PASJ}

\def\cm{\textrm{cm}}
\def\erg{\textrm{erg}}
\def\kpc{\textrm{kpc}}
\def\pc{\textrm{pc}}

\def\Kelv{\textrm{K}}
\def\sr{\textrm{sr}}
\def\ergps{\textrm{erg}~\textrm{s}^{-1}}
\def\kms{\textrm{km}~\textrm{s}^{-1}}
\def\eV{\textrm{eV}}

\def\GeV{\textrm{GeV}}

\def\PeV{\textrm{PeV}}
\def\EeV{\textrm{EeV}}
\def\yr{\textrm{yr}}
\def\Myr{\textrm{Myr}}
\def\Gyr{\textrm{Gyr}}
\def\muGauss{\mu\textrm{G}}
\def\Msun{\textrm{M}_{\sun}}
\def\Blue{}

\title[Fermi Bubbles as Wind Termination Shocks]{The Fermi Bubbles as Starburst Wind Termination Shocks}
\author[Lacki]{Brian C. Lacki$^{1,2}$\\$^1$Jansky Fellow of the National Radio Astronomy Observatory\\$^2$Institute for Advanced Study, Einstein Drive, Princeton, NJ 08540, USA, brianlacki@ias.edu}

\begin{document}
\maketitle

\begin{abstract}
The enhanced star formation in the inner 100 pc of the Galaxy launches a superwind {\Blue at} $\sim 1600\ \kms$ for M82-like parameters.  The ram pressure of the wind is very low compared to more powerful starburst winds.  I show that halo gas stops the wind a few kpc from the Galactic Centre.  I suggest that the termination shock accelerates cosmic rays, and that the resulting Inverse Compton $\gamma$-rays are visible as the Fermi Bubbles.  The Bubbles are then wind bubbles, which the starburst can inflate within 10 Myr.  They {\Blue can} remain in steady state as long as the starburst lasts.  The shock may accelerate PeV electrons and {\Blue EeV} protons.  The Bubbles may be analogues of galactic wind termination shocks in the intergalactic medium.  I discuss the advantages and problems of this model.  I note that any jets from Sgr A$^{\star}$ must burrow through the starburst wind bubble before reaching the halo gas, which could affect the early evolution of such jets.
\end{abstract}


\section{Introduction}
\label{sec:Introduction}
Powerful outflows are ever present in starburst regions ($\Sigma_{\rm SFR} \ga 0.1\ \Msun\ \yr^{-1}\ \kpc^{-2}$), blasting out at speeds $\gg 100\ \kms$ (\citealt{Chevalier85}{\Blue, hereafter} CC85; \citealt*{Heckman90}; \citealt{Strickland09}{\Blue, hereafter SH09}).  Winds erupt in these intense regions when superbubbles and supernova remnants overlap as the star-formation density increases \citep{McKee77}.  The hot plasma fills starbursts and blows out in a sound-crossing time, continuously replenished by massive stars.  Even in normal galaxies like the Milky Way, cosmic rays (CRs) can drive winds once they diffuse far enough from the galactic plane \citep*[][{\Blue hereafter B91}]{Breitschwerdt91}.  But what happens to the wind after it escapes these central regions?  

In the CC85 model, the wind rapidly accelerates as it passes the sonic point, reaching an asymptotic speed $v_{\Blue 1} \approx 1600\ \kms$ if supernova energy thermalizes efficiently and the mass loading is small{\Blue,} as in M82 {\Blue (SH09)}.  As the wind expands adiabatically out beyond the confines of the starburst, {\Blue random particle} motions in the wind slow down{\Blue in} both the wind plasma itself and CRs within the winds \citep*{Volk96}.  The random kinetic energy of the wind is converted into bulk kinetic energy, pushing the wind plasma like a piston.

Eventually, it becomes so rarefied that the ram pressure equals whatever external pressure there is.  At this location, a termination shock stops the wind.  These shocks are theorized to host energetic phenomena, such as {\Blue CR} acceleration.  {\Blue T}he termination shock of starbursts lie in the intergalactic medium {\Blue (IGM)}.

A prototype for starburst regions and outflows exists in our Galaxy, in the form of the Central Molecular Zone (CMZ).  Lying within 100 pc of Sgr A$^{\star}$, the region is filled with gas \citep{Molinari11}, a large star-formation rate density ($\Sigma_{\rm SFR} \approx 2\ \Msun\ \yr^{-1}\ \kpc^{-2}$; \citealt{YusefZadeh09}), strong magnetic fields \citep{Crocker10}, and TeV emission \citep{Aharonian06}.  Evidence for an outflow includes the abnormally weak nonthermal radio and $\gamma$-ray emission \citep{Crocker11-Wind}, suggesting that CRs are advected away from the CMZ \citep{Crocker11-Wild}.

The Fermi Bubbles are more evidence for powerful phenomena in the Galactic Centre (\citealt{Finkbeiner04,Dobler10}; \citealt*{Su10} {\Blue(S10)}; \citealt{Ade12}).  These bilobal structures, visible in $\gamma$-rays and radio, extend nearly 10 kpc from the Galactic Plane and emanate from the Centre.  Soft X-ray emission is suggestive of a shock at the Bubble edges \citep[][{\Blue S10}]{Sofue00,BlandHawthorn03}.  Most theoretical work interprets these bubbles as the result of an outburst from Sgr A$^{\star}$ (\citealt{Cheng11}; \citealt*{Zubovas11}; \citealt{Guo12}).  In these models, CR electrons ($e^-$) shine in radio by synchrotron emission and in $\gamma$-rays by Inverse Compton (IC) emission.  But, since these losses are rapid, either the $e^-$ are transported rapidly or they must be accelerated in place.

\citet{Crocker11-Bubbles} proposed {\Blue that} the CMZ powers the Bubbles \citep[see also][]{Crocker13}.  In this model, the CMZ has been accelerating CRs for possibly the Galaxy's {\Blue entire} history.  The wind advects CR protons a few kpc, where they accumulate for $10^8$--$10^{10}$ years.  The Bubbles' $\gamma$-ray emission is pionic; secondary $e^{\pm}$ made by pionic interactions emit the radio waves.  This model evades the severe adiabatic losses in a CC85-style wind by postulating a slow wind that becomes compressed in the Bubbles.  Leptonic starburst models are possible too: \citet{Biermann10} argued that {\Blue CR} $e^{\pm}$ rapidly diffuse from the {\Blue CMZ} and are advected into the Bubbles, where they emit IC.

\begin{figure}
\centerline{\includegraphics[width=8cm]{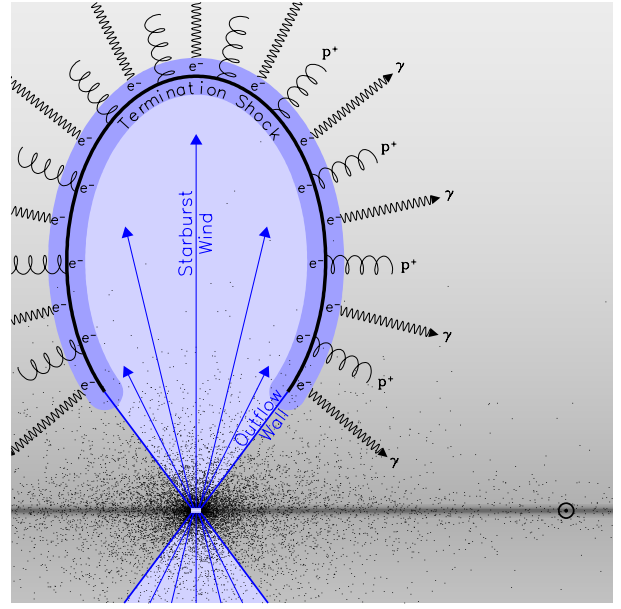}}
\caption{Sketch of the CMZ and its {\Blue wind} termination shock with respect to the Galaxy.  CR $e^-$ and $p^+$ are accelerated at the shock; {\Blue leptonic} emission appears as the radio and $\gamma$-ray bubbles.\label{fig:Geometry}}
\end{figure}

In this Letter, I propose that the Fermi Bubbles are the termination shocks of the Centre's starburst wind (Figure~\ref{fig:Geometry}).  Because the CMZ wind is less dense than prototypical starbursts' winds, its ram pressure is quite low.  Thus, the wind is stopped by the Galactic halo.  These shocks accelerate CRs{\Blue, particularly $e^-$ that radiate IC}.  Then, the Fermi Bubbles are a nearby analogue of the intergalactic termination shocks of more powerful starbursts \citep[c.f.,][]{Jokipii87,Volk04}.  

There is a second possibility: the Bubbles are powered by Sgr A$^{\star}$, but that the jet or outflow first ploughed through already-extant starburst wind bubbles.  This is different from current simulations of Sgr A$^{\star}$ outbursts, which assume that the gas in these regions was initially halo gas \citep[e.g.,][]{Guo12,Yang12}.  The effects of a starburst wind bubble on Sgr A$^{\star}$ outbursts may need to be explored.

\section{Properties of the CMZ Wind and its Termination Shock}
\subsection{Basic wind properties}
My discussion is informed by the CC85 theory of starburst winds, as amended by {\Blue SH09} to apply to disc geometries.  A hot plasma fills the wind volume, with properties determined by energy and mass injection.  The CMZ is a disc forming stars at a rate ${\rm SFR} \approx 0.1\ \Msun\ \yr^{-1}$ {\Blue \citep{YusefZadeh09,Immer12}}.  {\Blue TeV $\gamma$-ray emission, assumed by \citet{Crocker11-Wild} to trace star formation, is detected within $|\ell| < 0.8^{\circ}$ and $|b| < 0.3^{\circ}$ \citep{Aharonian06}, indicating a radius $R_{\rm CMZ} = 112\ \pc$ and midplane-to-edge height $h_{\rm CMZ} = 42\ \pc$ for a distance of 8.0 kpc \citep[c.f.][]{Molinari11}.}  Mass is injected into the wind at a rate $\dot{M} = 0.117 \beta \times {\rm SFR}$ from stellar winds and explosions, and these phenomena heat the wind at a rate $\dot{E} = 2.54 \times 10^{41} \epsilon_{\rm therm} [{\rm SFR} / (\Msun\ \yr^{-1})]\ \erg\ \sec^{-1} $.\footnote{I have converted the values in {\Blue SH09} to a Salpeter IMF running from 0.1$\ \Msun$ to 100$\ \Msun$.}   I take $\beta = 2$ and $\epsilon_{\rm therm} = 0.75$, as for M82 {\Blue (SH09)}.  Then the central electron density of the wind is given by
\begin{align}
n_c = \frac{\displaystyle 0.518 (\mu m_H)^{-1} \sqrt{\dot{M}^3 / \dot{E}}}{R_{\rm CMZ} (R_{\rm CMZ} + 2 h_{\rm CMZ})} = 0.015\ \cm^{-3} \times {\rm SFR}_{0.1}
\end{align}
{\Blue with a} central temperature {\Blue of}
\begin{align}
T_c = (\gamma - 1) \mu m_H \dot{E} (\gamma k_B \dot{M})^{-1} = 3.7 \times 10^7\ \Kelv,
\end{align}
for a mean molecular weight $\mu = 0.59${\Blue,} adiabatic index $\gamma = 5/3${\Blue, and star-formation rate ${\rm SFR} = {\rm SFR}_{0.1} \times (0.1\ \Msun\ \yr^{-1})$}.  I note that there is evidence for diffuse X-ray plasma with $n_e \approx 0.05\ \cm^{-3}$ and $T \approx 8 \times 10^7\ \Kelv$ \citep{Uchiyama13}; the order-of-magnitude agreement {\Blue suggests} that {\Blue it} is the CC85 wind plasma, but the reason for the {\Blue density} discrepancy is unknown.

In the CC85 model, the wind flows out of the starburst at a speed $v_{\Blue 1} = \sqrt{2 \dot{E} / \dot{M}} = 1600\ \kms$.  The escape speed from the CMZ is only 800 -- $1000\ \kms$ ({\Blue B91}; \citealt*{Launhardt02}), so the wind {\Blue slows down to $v_{\Blue 2} = 1300\ \kms$}.\footnote{\citet{Carretti13} argue that the radio ``spurs'' trace CMZ features if the outflow has a speed of $1000\ \kms$.}  Although CC85 assume spherical symmetry, actual starburst winds are observed to flow in cones {\Blue \citep[c.f.][]{BlandHawthorn03}}.  I therefore assume the outflow cone{\Blue s} fill a solid angle $\Omega$.  The density of the outflow is $\rho = \dot{M} / (\Omega v_{\Blue 2} r^2)$, and the ram pressure is $P_{\rm ram} = \rho v_{\Blue 2}^2 / 2$. 

\subsection{The termination shock}
The termination shock occurs when the ram pressure is equal to the external pressure {\Blue $P_{\rm ext}:$}
\begin{equation}
R_t = \sqrt{\frac{\dot{M} v_{\Blue 2}}{2 \Omega {\Blue P_{\rm ext}}}}.
\end{equation}
The state of the gas several kpc into the Galactic halo is poorly known.  I take $P_{\rm ext} / k_B = 1000\ \Kelv\ \cm^{-3}$, consistent with loose constraints from Galactic absorption line features (e.g., \citealt{Savage03}{\Blue, hereafter S03}; \citealt{Yao05,Hsu11}) and hydrodynamics (e.g., \citealt{Spitzer56,Wolfire95}; \citealt*{Fang13}).

For the CMZ wind, 
\begin{align}
R_t = {\Blue 4.9}\ \kpc \left[{\rm SFR}_{0.1} \left(\frac{\Omega}{\pi} \frac{P_{\rm ext} / k_B}{10^3\ \Kelv\ \cm^{-3}}\right)^{-1}\right]^{1/2},
\end{align}
roughly the size of the Fermi Bubbles.  For SFRs of ${\Blue \ga 1}\ \Msun\ \yr^{-1}$, the termination shock is instead firmly in the realm of circumgalactic or even intergalactic distances.

The termination shock is a steady-state feature, present even if the starburst wind has always been active.  The wind just inside the shock is in pressure equilibrium with the gas outside.  In contrast, many models of the Fermi Bubbles propose that the shock is transient, resulting from a recent burst of energy injection, and the shock is driven outwards by an overpressure.

\subsection{Evolution of the wind bubbles}
Some authors previously proposed a {\Blue transient} wind from a brief CMZ starburst 15 Myr ago; the shock front from the wind could appear as radio/X-ray bubbles \citep{Sofue00,BlandHawthorn03}. If the starburst wind just turned on, how long does it take to reach steady state and inflate the bubbles?  Suppose that the medium surrounding the CMZ is uniform, with density $\rho_{\rm ext}$ and pressure $P_{\rm ext} = \rho_{\rm ext} \sigma^2 / 2$.  The wind turns on, and starts sweeping up this medium at a shock.  Once it reaches a radius $R_{\rm Sed} \approx \sqrt{3 \dot{M} / (\Omega v_{\Blue 2} \rho_{\rm ext})}$, the mass of swept up material is greater than the mass in the wind, and the shock slows down -- much like a supernova remnant does, except there is continuous mass injection.  As long as $\sqrt{3} \sigma < v_{\Blue 2}$, $R_{\rm Sed} < R_t$; for the conditions I have been assuming, $R_{\rm Sed} \approx 1\ \kpc$.  

After this point, I estimate the shock speed by assuming that the bubbles are adiabatic, and that the {\Blue bulk} kinetic energy of the swept up material is a fraction $\kappa$ of the injected mechanical power{\Blue, with the rest converting into heat}.  The {\Blue expansion speed} of the shock is $dR_s/dt = \sqrt{6 \kappa \dot{E} t / (\rho_{\rm ext} \Omega R_s^3)}$, giving $t = R^{5/3} [(3/50) \rho_{\rm ext} / (\dot{E} \kappa)]^{1/3}$.  It takes 
\begin{equation}
t_{\rm inflate} = \left[\frac{9}{80000} \frac{\dot{M}^5 v_{\Blue 2}^5 \rho_{\rm ext}^2}{\kappa^2 \dot{E}^2 \Omega^3 P_{\rm ext}^5}\right]^{1/6} \approx {\Blue 8} \kappa^{-1/3}\ \Myr
\end{equation}
for the shock to reach {\Blue $R_t$} (with $\rho_{\rm ext} = 0.001 m_H\ \cm^{-3}$ and $P_{\rm ext} = 1000\ \Kelv\ \cm^{-3}$), where the shock expands at a speed $v_s = 0.761 \kappa^{1/3} (v_{\Blue 2}/\sigma)^{1/3} \sigma \approx {\Blue 210} \kappa^{1/3}\ \kms$.  

{\Blue What happens after the shock overshoots $R_t$ depends on whether the halo gas is replenished.  If the halo gas is static, the bubbles expand as the CMZ continue to inject power.  The post-shock gas only has enough energy to reach a distance of 50 kpc before it falls back in (B91), even with no radiative losses.  Since the swept-up halo and post-shock wind is denser than the free-expanding wind, Rayleigh-Taylor instabilities develop, with fingers of infalling gas pushing back down through the outflowing wind \citep{King10}.  However, if the halo gas is actually an IGM headwind as the Galaxy moves or a Galaxy-scale outflow (e.g., S03, \citealt{Everett08}), then a bow shock forms around the Bubbles, analogous to stellar wind bow shocks.  The post-shocked gas is swept around the Bubbles edges \citep[e.g.,][]{Wilkin96} and eventually mixes with the larger-scale flow.  It may then fall back into the Galaxy as a fountain (S03).  The termination shock itself oscillates and settles at $R_t$.}


In my model, $t_{\rm inflate}$ is longer than the minimum time $H / \dot{E} \approx 2\ \Myr$ for the starburst to do enough work to inflate the Fermi Bubbles, where $H = \gamma P_{\rm ext} [\Omega R_t^3 / 3]$ is the enthalpy \citep[c.f.,][{\Blue hereafter C12}]{Crocker12}.  The reason is that the swept up material is not simply pushed out of the bubbles, but accelerated, requiring more energy.  On the other hand, my estimate of $H / \dot{E}$ is $\sim 3000$ times lower than the $10\ \Gyr$ estimated by {\Blue C12}.  The major discrepancy {\Blue results from the somewhat large (and more accurate) Bubble volume, far greater halo pressure ($4 \times 10^4\ \Kelv\ \cm^{-3}$), and weaker outflow power assumed in C12.}  In any case, the CMZ {\Blue clearly} can inflate the Fermi Bubbles.

What if the wind suddenly shuts off?  Then the bubbles collapse at a speed $\sim \sigma$, specifically the {\Blue halo} sound speed.  The collapse time is very long, though, almost 40 Myr.  The large stellar \citep{Launhardt02} and gas \citep{Molinari11} masses in the nuclear bulge are consistent with star-formation over several Gyr.  Even if the output power of the CMZ fluctuates on time-scales of $\la 10\ \Myr$, as extragalactic true nuclear starbursts do \citep{Mayya04}, the termination shock remains roughly in the same place.  Thus, the Fermi Bubbles may be a relatively permanent feature of our Galaxy, as \citet{Crocker11-Bubbles} originally argued.  

\section{Nonthermal Emission from the Shock}
Can the radiation observed from the Fermi Bubbles be identified with the starburst wind termination shocks? 

Because of adiabatic losses, only a small fraction of CR energy density remains as the wind reaches {\Blue $R_t$}, regardless of any additional radiative losses.  Yet neither the CR nor the thermal energy disappears during adiabatic expansion; it simply converts into bulk kinetic energy.  At the termination shock, that energy is converted back into random particle energy.  Most of the energy goes into heating the gas, but an appreciable factor ($\ga 10\%$) is {\Blue expected to be} in the form of {\Blue CRs} accelerated at the shock {\Blue \citep[e.g.,][]{Morlino12,Kang13,Caprioli14}}. Thus, termination shocks at the Bubble edges inject relativistic particles \emph{in situ}, circumventing any losses.

\subsection{Estimate of $\gamma$-ray emission}
The mechanical power of the wind is just $\dot{E}$.  Suppose the shocks convert $\eta = 30\%$ of this power into CR protons, which are accelerated with a $E^{-2}$ spectrum.  I define $\Psi = \ln(E_{\rm max} / E_{\rm min})$, where $E_{\rm min}$ and $E_{\rm max}$ are the minimum and maximum energies of CR protons in the spectrum.  Then the injection spectrum of CR protons is:
\begin{equation}
E^2 \frac{dQ_p}{dE} \approx {\Blue 2.0} \times 10^{38}\ \ergps\ {\rm SFR}_{0.1} \eta_{0.3} \Psi_{20}^{-1},
\end{equation}
where $\eta_{0.3} = \eta / 0.3$ and $\Psi_{20} = \Psi / 20$.  The protons presumably {\Blue escape the Galaxy} without radiating.

Collisionless shocks also accelerate primary CR {\Blue $e^-$}, but $\ga \GeV$ {\Blue $e^-$} are thought to carry little of the energy.  In the test particle approximation to diffusive shock acceleration, at energies $\ga m_p c^2$, the ratio of injected {\Blue $e^-$} to injected protons is $\tilde\delta = (m_p / m_e)^{(p-1)/2}$ \citep{Bell78}.  Thus, for $p = 2.0$, we have $\tilde\delta \approx 43$:
\begin{equation}
E^2 \frac{dQ_e}{dE} \approx {\Blue 6.7} \times 10^{36}\ \ergps\ {\rm SFR}_{0.1} \eta_{0.3} \Psi_{20}^{-1}.
\end{equation}

Radiative losses for electrons are fast near the Galaxy, and the IC cooling time for an {\Blue $e^-$} is
\begin{equation}
t_{\rm IC} = 3.1\ \Myr \left(\frac{E}{100\ \GeV}\right)^{-1} \left(\frac{U_{\rm rad}}{1\ \eV\ \cm^{-3}}\right)^{-1}.
\end{equation}
I assume that a fraction $f_{\rm IC} \approx 1$ of the power goes into IC, with the rest mostly going into synchrotron.  In the Thomson limit, the energy of an IC upscattered photon goes as $E_{\gamma} \propto E_e^2$, where $E_e$ is the relativistic electron energy.  Thus, each dex of electron energy is stretched out into 2 dex of $\gamma$-ray energy, and we have $E^2 dL_{\rm \Blue IC}/dE \approx (f_{\rm IC} / 2) E^2 dQ_e/dE$.  The average distance of the Bubbles is 10 kpc, and they cover 0.8 sr {\Blue (S10)}.  So the flux from the Fermi Bubbles at Earth is
\begin{equation}
E^2 \frac{dN_{\rm IC}}{dE} \approx {\Blue 150}\ \eV\ \cm^{-2}\ \sec^{-1}\ \sr^{-1}\ f_{\rm IC} {\rm SFR}_{0.1} \eta_{0.3} \Psi_{20}^{-1}.
\end{equation}
This {\Blue is about half of} the $\gamma$-ray flux from the Fermi Bubbles {\Blue (S10)}; {\Blue higher fluxes result if the CMZ SFR was higher in the past, or if $\Psi < 20$ (with fewer high energy particles).}

\subsection{Magnetic fields}
Synchrotron microwaves from the Bubbles demonstrate that magnetic fields are present.  Its 23 GHz brightness temperature is $\sim 50\ \mu{\rm K}$ at $|b| \la 30^{\circ}$ \citep{Dobler12-WMAP,Ade12}.  From minimum energy arguments \citep{Beck05} and $\gamma$-ray to radio ratios \citep{Dobler12-B,Hooper13}, the magnetic fields are roughly 4 -- 6 $\muGauss$.  The radio ``spurs'' may have higher magnetic fields, $\sim 15\ \muGauss$ \citep{Jones12}.  At higher latitudes, the microwave emission vanishes, indicating low $B$ \citep{Dobler12-WMAP,Ade12}. 

These magnetic fields pose two challenges to my model.  First, the synchrotron losses cannot be much faster than IC, or else there is not enough wind power to explain the $\gamma$-ray emission.  But for $B \la 6\ \muGauss$, $f_{\rm IC} \ga 0.5$.  

Second, if the magnetic pressure is much greater than $P_{\rm ext}$, that could invalidate my calculations of $R_{\Blue t}$.  For $B = 4\ \muGauss$, the magnetic pressure is $5000\ \Kelv\ \cm^{-3}$, with a similar pressure in CRs.  But the wind ram pressure at 4 kpc from the CMZ is only $2000\ \Kelv\ \cm^{-3}$.  The tension is eased if the CMZ SFR was $\sim 0.2\ \Msun\ \yr^{-1}$ in the recent past.

\subsection{Ultra-high energy CRs from the shocks?}
The maximum CR energy is limited by the magnetic field in the shock, the size of the acceleration region, and the radiative losses the CRs experience.  The sheer size of the Bubbles makes them excellent places for the acceleration of extremely high energy particles \citep[c.f.,][]{Cheng12}.

Electrons are highly radiative particles, so their maximum energy is set by IC or synchrotron losses.  They gain a fraction $\sim v_{\Blue 2} / c$ of their energy per Larmor orbit, for an acceleration time of $t_{\rm acc} \approx 20 E / (3 v_{\Blue 2} e B)$ that rises with energy \citep{Gaisser90}.  By setting $t_{\rm acc}$ to the energy loss time from synchrotron and IC losses, I derive
\begin{equation}
E_e \la \sqrt{\frac{9 e B v_{\Blue 2}}{80 \sigma_T U c}} m_e c^2 \approx {\Blue 0.5}\ \PeV \left(\frac{B}{5\ \muGauss}\right)^{1/2} \left(\frac{U}{\eV\ \cm^{-3}}\right)^{-1/2}.
\end{equation}
$U$ includes the sum of the CMB energy densities and the magnetic field energy densities; Klein-Nishina effects suppress losses off starlight for PeV {\Blue $e^-$}, but {\Blue they} still {\Blue upscatter} the CMB.  These $\gamma$-rays cascade down to TeV energies due to $\gamma\gamma$ absorption within the Galaxy \citep*{Moskalenko06}.  TeV and PeV {\Blue $e^-$} are a possible source of TeV $\gamma$-rays from the Fermi Bubbles.

For protons, the most important constraint on the maximum energy is that their Larmor radius cannot be larger than the acceleration region \citep{Hillas84}.  Requiring that $E_p \la R_{\rm max} Z e B$, I find
\begin{equation}
E_p \la 5 Z\ \EeV \left(\frac{R_{\rm max}}{1\ \kpc}\right) \left(\frac{B}{5\ \muGauss}\right).
\end{equation}
It takes $t_{\rm acc} \approx {\Blue 5}\ \Myr$ to reach these energies if $B = 5\ \muGauss$.

Nuclei at these energies are subject to photopion and {\Blue pair production} losses off the far infrared radiation from the Milky Way.  {\Blue P}hotopionic $\gamma$-rays cascade down to $\sim$100 TeV as they propagate to Earth.  Photopions can also decay into neutrinos, which could be detected at $\sim$10 PeV.  

\section{Discussion}

The starburst power source of this model is known, the wind's existence is fairly well established, and its termination shock is almost inevitable if the wind becomes supersonic.  Because the injection sites are the termination shocks, there are no problems {\Blue from} cooling.  IC emission does not require dense reservoirs of gas or long time-scales.  Finally, the shock could be a source of TeV IC emission, PeV neutrinos, and EeV CRs.  

There are several problems that must be addressed.  First, why does the $e^-$ spectrum harden at low energies ({\Blue S10}; \citealt{Ade12})?  This could be a cooling break, but only if the electrons are just a few Myr old.  Yet the CMZ is likely older ($\gg 10\ \Myr$), so older electrons are probably present on some level.  Have the older electrons simply diffused away?  Perhaps the CMZ fluctuates on time-scales of $\sim 10\ \Myr$, but is steady on longer time-scales {\Blue \citep[c.f.,][]{Kruijssen14}}.  Then, the Fermi Bubbles' IC emission represents the {\Blue $e^-$} from the most recent pulse of star formation.  Either way, low energy IC and radio emission should be visible surrounding the Fermi Bubbles.  Or is the break in the injection spectrum itself?  

Second, why are the Bubbles elongated?  This indicates the termination shock radius depends on angle, which could happen if the external pressure depends strongly on midplane distance instead of distance to the Galactic Centre, or if power is concentrated along the axis of the outflow cone{\Blue s}.

Third, this model cannot explain the possible $\gamma$-ray jets discovered by \citet{Su12} in the inner regions of the Bubbles, which could be from Sgr A$^{\star}$ itself.  The ``jets'' could instead be related to stellar clusters in the CMZ \citep[c.f.,][]{Carretti13}; perhaps they are smaller cluster winds within the CMZ wind.  

Finally, the flat $\gamma$-ray surface brightness of the Bubbles is inconsistent with IC emission from a thin shell around the shocks {\Blue (S10)}.  If there is turbulence within the wind bubble, second order acceleration might be able to solve this problem, but the model proposed by \citet{Mertsch11} proposes that the shock is growing instead of steady-state; any turbulence must be generated before the shock itself.  Alternatively, if the wind fluctuates on Myr time-scales, internal shocks may develop and accelerate CRs within the Bubbles \citep[c.f.,][]{Dorfi12}.  {\Blue Finally, Rayleigh-Taylor instabilities at the shocks can develop as the extremely rarefied wind interacts with the relatively dense halo gas \citep{King10}.  These instabilities corrugate the shock, mixing halo gas into the interior of the Bubbles and leading to CR acceleration inside the Bubbles themselves \citep{Zubovas11}.}  

On the other hand, even if activity in Sgr A$^{\star}$ itself powers the Bubbles, the older CMZ starburst still likely inflated wind bubbles of some kind.  The jet or wind launched from Sgr A$^{\star}$ first traversed these older bubbles.  This is a different scenario from those in simulations, where the Sgr A$^{\star}$ outflow entered into the hydrostatically-supported halo directly.  The effects of starburst wind bubbles on the evolution of such outflows should be studied. 

\section*{Acknowledgements}
I thank Roland Crocker{\Blue,} Todd Thompson{\Blue, and the referee} for their comments.  I was supported by a Jansky Fellowship from the National Radio Astronomy Observatory.  The National Radio Astronomy Observatory is operated by Associated Universities, Inc., under cooperative agreement with the National Science Foundation.

\end{document}